\documentclass{llncs}
\usepackage{amsmath,amsfonts,amssymb}
\usepackage{graphicx}
\usepackage{algorithm}
\usepackage{algorithmic}
\def\B#1{\mathbf{#1}}

\begin{document}

\title{Detection of brain functional-connectivity difference in
post-stroke patients using group-level covariance modeling}

\author{Ga\"el Varoquaux\inst{1}, Flore Baronnet\inst{2},
Andreas Kleinschmidt\inst{2}, Pierre Fillard\inst{1}
and Bertrand Thirion\inst{1}}

\institute{Parietal project-team, INRIA Saclay-\^ile de France
\and
INSERM, U562, CEA/Neurospin b\^at 145, 91191 Gif-Sur-Yvette, 
France}

\maketitle

\begin{abstract}
\vspace*{-1ex}%
Functional brain connectivity, as revealed through distant correlations
in the signals measured by functional Magnetic Resonance Imaging (fMRI),
is a promising source of biomarkers of brain pathologies. However,
establishing and using diagnostic markers requires probabilistic
inter-subject comparisons. Principled comparison of
functional-connectivity structures is still a challenging issue.
We give a new matrix-variate probabilistic model suitable for
inter-subject comparison of functional connectivity matrices on the
manifold of Symmetric Positive Definite (SPD) matrices. We show that this
model leads to a new algorithm for principled comparison of connectivity
coefficients between pairs of regions.
We apply this model to comparing separately post-stroke patients to a
group of healthy controls. We find neurologically-relevant connection
differences and show that our model is more sensitive that the standard
procedure. To the best of our knowledge, these results are the first
report of functional connectivity differences between a single-patient
and a group and thus establish an important step toward using functional
connectivity as a diagnostic tool.
\vspace*{-1ex}%
\end{abstract}

\section{Introduction}
\vspace*{-1ex}%

The correlation structure of brain activity, measured via fMRI, reveals
stable inter-subject networks of distant brain regions that can be the
expression of cognitive function. 
In particular, some connectivity networks are present in the absence of
stimuli. They can reveal intrinsic brain activity and are studied in the
{\sl resting-state} paradigm. These structures are of particular interest
to study and diagnose brain diseases and disorders \cite{greicius2008b}
as they can be used for deep probes of brain function on diminished
subjects. Not only can they extract medically or cognitively relevant
markers on subjects unconscious \cite{vanhaudenhuyse2010}, or with
limited cooperation \cite{fair2007}, but they also give information on
higher-level cognitive systems that are challenging to probe via medical
imaging or behavioral clinical tests \cite{uddin2009}.

To use functional connectivity as a quantitative inference tool,
principled probabilistic comparison of connectivity structures across
subjects is needed. Unlike with stimuli-response studies used routinely
in functional neuroimaging, this comparison is challenging, as the
underlying description of the signal is multivariate: each brain-activity
time course is considered relative to others. Univariate group models,
such as random effects or mixed effects, are in general not sound as they
neglect the strong statistical dependence between parameters estimated
from the data. Multivariate techniques have been successfully employed to
single out outlying subjects \cite{sato2008}, but have met little
success: their results are difficult to interpret as they do not point to
specific localized differences.


In this paper, we focus on the description of brain
functional-connectivity using inter-regions correlation matrices. We
first review the current practice in inter-subject functional covariance
comparison and recall some results on the manifold of covariance
matrices. Then, we introduce a probabilistic model at the group level for
the different subjects' correlation matrices, and a corresponding
algorithm to detect connectivity differences in a specific
parametrization of the covariance matrices, as correlations are a form of
covariance. We quantify on simulated data the performance of this
detection. Finally, we apply the model to the individual comparison of
the connectivity structure of stroke patients to a group of healthy
controls, and show that it outperforms the current practice.

\section{State of the art}
\vspace*{-1ex}%

\subsection{Problem statement: comparing functional brain connectivity}

We consider $S$ subjects, represented by the correlation
matrices between brain-activity time series extracted from $n$ ROIs:
$\{\B{\Sigma}^s \in\mathbb{R}^{n \times n}, s = 1 \dots S\}$. The
challenge is to give a probabilistic description of the population of
correlation matrices so as to find the significant differences between
subjects or groups.
The current practice in functional-connectivity studies is to compare the
coefficients of the correlation matrices across subjects (see for
instance \cite{fair2007,rocca2007}). This procedure can be expressed as
a univariate additive linear model on the correlation matrix:
\begin{equation}
    \B{\Sigma}^s = \B{\Sigma}^\star + \B{d\Sigma}^s
    \label{eqn:flat_model}
\end{equation}
where $\B{\Sigma}^\star$ is a covariance matrix representative of the
mean effect, or the group, and $\B{d\Sigma}^s$ encode subject-specific
contributions.

However, with this description it is difficult to isolate significant
contributions to $\B{d\Sigma}^s$. Indeed, for interpretation, some
coefficients are zeroed out, eg by thresholding a test statistic, as in
\cite{fair2007}, which eventually leads to a non positive definite
matrix, for which it is impossible to write a multivariate normal
likelihood. As a result, the subject-variability description learned on a
population cannot give probabilistic tests on new subjects.

\subsection{Recent developments on the covariance-matrix manifold}


The mathematical difficulty stems from the fact that the space of SPD
matrices, $\mathcal{S}ym_n^+$, does not form a vector space: $\B{A}, \B{B}
\in \mathcal{S}ym_n^+ \nRightarrow \B{A} - \B{B} \in \mathcal{S}ym_n^+$.
The Fisher information matrix of the multivariate normal
distribution can be used to construct a metric on a parametrization of
covariance matrices \cite{lenglet2006} and thus define $\mathcal{S}ym_n^+$
as a Riemannian
manifold that is well-suited for performing statistics on
covariances \cite{pennec2006}.
Local differences on this manifold can be approximated by vectors of the
tangent space: 
if $\B{B}$ is close enough to $\B{A}$, 
the application:
\mbox{
$
    \phi_\B{A}: \B{B} \rightarrow
    \log 
    \bigl( \B{A}^{-\frac{1}{2}}\B{B}\B{A}^{-\frac{1}{2}} \bigr)
$
}
maps locally the bipoint $ \B{A}, \B{B} \in \mathcal{S}ym_n^+ \times
\mathcal{S}ym_n^+$ to $
\overrightarrow{\B{A}\B{B}} \in \mathcal{S}ym_n$, the space of symmetric
matrices\footnote{Note that we do not use the same definition of the
mapping as in \cite{pennec2006,lenglet2006}, as we are interested in
mapping to $\mathcal{S}ym_n$, the tangent space around $\B{I}_n$, and not
the tangent space in $\B{A}$. It extracts a
statistically independent parametrization
(Eq. (\ref{eqn:exact_tangent_model}) and (\ref{eqn:tangent_gaussian})).}.
A convenient parametrization of $\overrightarrow{\B{W}} 
\in \mathcal{S}ym_n$ is
$\text{Vec}(\B{W}) = \{\sqrt{2}\, w_{i, j}, \, j<i \,,\, w_{i, i}, i =
1\dots n\}$ that forms an orthonormal basis of the tangent space
\cite{pennec2006}. Finally,
$
    \bigl\| \overrightarrow{\B{A}\B{B}} \bigr\|^2_{\B{A}}
	    = \bigl\| \log \bigl(
		\B{A}^{-\frac{1}{2}} \B{B} \B{A}^{-\frac{1}{2}} 
	      \bigr) \bigr\|^2_2
$
gives the intrinsic norm of $\overrightarrow{\B{A}\B{B}}$
on the $\mathcal{S}ym_n^+$ manifold, according to the metric
around point $\B{A}$: the distance between $\B{A}$ and $\B{B}$ in the
manifold.

\section{Matrix-variate random effects model for covariances}
\vspace*{-1ex}%

\label{sec:matrix_variate_model}

\paragraph{Multi-subject probability distribution for covariance matrices}

Using the Riemannian parametrization of $\mathcal{S}ym_n^+$, we describe
the individual correlation matrix population as a distribution of
matrices scattered around a covariance matrix representative of the
group, $\B{\Sigma}^{\star}$. As this distribution must be estimated with
a small number of observations compared to the feature space, we model it
using the probability density function that minimizes the information
with a given mean on the manifold, the generalized Gaussian
distribution \cite{pennec2006}:
\begin{equation}
    p(\B{\Sigma}) = k(\sigma) \exp \biggl(-\frac{1}{2\sigma^2}  
	\bigl\| \overrightarrow{\B{\Sigma}^{\star}\B{\Sigma}}
	\bigr\|^2_{\B{\Sigma}^{\star}} 
    \biggr),
    \label{eqn:gaussian}
\end{equation}
where $\sigma$ encodes an isotropic variance on the manifold and $k$ is a
normalization factor. Given multiple observations of $\B{\Sigma}$
corresponding to individual correlation matrices, $\B{\Sigma}^s$, the
maximum likelihood estimate of $\B{\Sigma}^{\star}$ is independent of
$\sigma$ and given by the Fr\'echet mean of the observations
\cite{pennec2006}, minimizing $\displaystyle\sum_s \bigl\|
\overrightarrow{\B{\Sigma}^{\star}\B{\Sigma}^s}
\bigr\|^2_{\B{\Sigma}^{\star}}$.

\paragraph{Parametrization in the tangent space}

We express the individual covariance matrices as
a perturbation of the group covariance matrix $\B{\Sigma}^\star$:
\begin{align}
\forall s=1\dots S, & &
    \B{\Sigma}^s = 
&
    \; \phi^{-1}_{\B{\Sigma}^\star} (\B{d\Sigma}^s)
    =
    \; {\B{\Sigma}^{\star}}^{\frac{1}{2}}\, \exp(\B{d\Sigma}^s)\,
    {\B{\Sigma}^{\star}}^{\frac{1}{2}},
    \quad
    \label{eqn:exact_tangent_model}
\\
\text{thus, using (\ref{eqn:gaussian}),}
& &
    p(\B{d\Sigma}^s) =
&
    \; k'(\sigma) \exp \biggl(-\frac{1}{2 \sigma^2}
	\bigl\| 
	    \B{d\Sigma}^s
	\bigr\|^2_2
	\biggr).
    \label{eqn:tangent_gaussian}
\end{align}
The parameters of $\text{Vec}(\B{d\Sigma}^s)$ follow a normal
distribution, with diagonal
covariance $\sigma$, and the maximum-likelihood estimate of $\sigma$ is
given by $\widehat{\sigma}_\text{MLE}^2 = \frac{1}{S}\sum_s \|
\text{Vec}(\B{d\Sigma}^s)\|_2^2$. The model can thus be interpreted as a
random-effect model on the parametrization of
$\text{Vec}(\B{d\Sigma}^s)$, in the space tangent to the manifold
$\mathcal{S}ym_n^+$.
Assuming that the distribution is narrow on the manifold, 
$\|\B{d\Sigma}^s\|_2 \ll 1$, eq. (\ref{eqn:exact_tangent_model}) can be seen as 
the application of the placement function to move a noise $\B{d\Sigma}^s$
isotropic around $\B{I}_n$ to $\B{\Sigma}^\star$ (see \cite{pennec2006},
section 3.5):
\begin{equation}
    \B{\Sigma}^s \simeq
    \; {\B{\Sigma}^{\star}}^{\frac{1}{2}}\, 
    (\B{I}_n + \B{d\Sigma}^s)\,
    {\B{\Sigma}^{\star}}^{\frac{1}{2}}.
    \label{eqn:tangent_model}
\end{equation}

\paragraph{Model estimation from the data}

We start from individual time-series of brain activity in selected
regions of interest, $\B{X} \in \mathbb{R}^{n\times t}$. We use the
Ledoit-Wolf shrinkage covariance estimator \cite{ledoit2004} for a good
bias-variance compromise 
when
estimating correlation matrices from $t$ time points with $n < t < n^2$.
From this estimate of individual correlation matrices, we compute the
intrinsic mean on $\mathcal{S}ym_n^+$ using algorithm 3 of
\cite{fletcher2007}. Finally, we estimate $\sigma$ from the residuals of
individual correlation matrices in the space tangent in
$\B{\Sigma}^\star$ (see algorithm \ref{alg:group_model}).

\begin{algorithm}[tb]
   \caption{Estimation of the group model}
   \label{alg:group_model}
\begin{algorithmic}[1]
   \STATE {\bfseries Input:} individual time series
$\B{X}^1 \dots \B{X}^s$.
   \STATE {\bfseries Output:} estimated group covariance matrix
$\widehat{\B{\Sigma}^\star}$,
group variance $\widehat{\sigma}$.
   \FOR{$s=1$ {\bfseries to} $S$}
	\STATE Compute 
	    $\widehat{\B{\Sigma}^s} \leftarrow \text{LedoitWolf}(\B{X}^s)$.
   \ENDFOR
   \STATE Compute $\widehat{\B{\Sigma}^\star} \leftarrow \text{intrinsic mean} 
	(\widehat{\B{\Sigma}^1} \dots \widehat{\B{\Sigma}^s})$.
   \FOR{$s=1$ {\bfseries to} $S$}
	\STATE Compute 
	    $\widehat{\B{d\Sigma}}^s \leftarrow 
		\widehat{\B{\Sigma}^\star}^{-\frac{1}{2}}
		\B{\Sigma}^s
		\widehat{\B{\Sigma}^\star}^{-\frac{1}{2}} - \B{I}_n 
	    $.
   \ENDFOR
   \vspace*{-.5ex}
   \STATE Compute $\widehat{\sigma} \leftarrow \sqrt{\frac{1}{S} 
\displaystyle\sum_s \bigl\|
\text{Vec}(\widehat{\B{d\Sigma}^s}) \bigl\|_2^2}$.
\end{algorithmic}
\end{algorithm}

\section{Testing pair-wise correlations statistics}
\vspace*{-1ex}%

The multivariate probabilistic model for correlations between regions
exposed above enables us to define an average correlation matrix of a
group, as well as the dispersion of the group in the covariance matrix
space. Thus it can be used to test if a subject is different to a control
group. However, to aid diagnosis, it is paramount to pin-point why such a
subject may be different.
In the tangent space, the parameters $d\Sigma_{i, j}^s$ of
$\text{Vec}(\B{d\Sigma}^s)$ are mutually independent. We can thus conduct
univariate analysis on these parameters to test which significantly
differs from the control group. However, the independence of the
parameters is true only in the space tangent at the population average
$\B{\Sigma}^\star$, of which we only have an estimate
$\widehat{\B{\Sigma}^\star}$. Thus, to account for projection error, we
resort to non-parametric sampling of the control population to define a
null distribution for the parameters $d\Sigma_{i, j}^s$.

Specifically, we are interested in testing if a difference observed for a
subject in one of the $d\Sigma_{i, j}^s$ can be explained by variation of
the control population. As the control population is typically small, we
generate the null distribution by leave one out: for each control
subject, we generate surrogate control populations $\tilde{S}$ by
bootstrap from the other control subjects and estimate the corresponding
average covariance $\widetilde{\B{\Sigma}^\star}$. We use
$\widetilde{\B{\Sigma}^\star}$ to project all the individual correlation
matrices, including the left out subject, to compute
$\widetilde{d\Sigma_{i, j}^s}$, and we do a one sample T test of the
difference between $\widetilde{d\Sigma_{i, j}^s}$ for the left out
subject with regards to the resampled control group $\tilde{S}$. We
record the values of this T test as an estimate of the null distribution
$P^o_{i,j}$ of the T test on the corresponding coefficient between the
controls and a patient. Finally, we estimate the average covariance
$\widehat{\B{\Sigma}^\star}$ for the complete group of controls and, for
each $k$ patient to investigate, we perform a T test of the difference
between $\widehat{d\Sigma_{i, j}^s}$ for the patient and the control
group. We use $P^o_{i,j}$ to associate a p-value to each coefficient
per subject. We correct for multiple comparisons using 
Bonferroni correction: for each patient, $\frac{1}{2}n\,(n-1)$ tests are
performed. The procedure is detailed in algorithm
\ref{alg:coefficient-level}.

\begin{algorithm}[tb]
   \caption{Coefficient-level tests}
   \label{alg:coefficient-level}
\begin{algorithmic}[1]
   \STATE {\bfseries Input:} individual time series for controls
$\B{X}^1 \dots \B{X}^s$ and a patient $\B{X}^k$, p-value $p$, number of
bootstraps, $m$.
   \STATE {\bfseries Output:} Pair-wise p-values $p_{i, j}$ controlling for 
    the difference in $d\Sigma_{i, j}$ between the patient and the control 
    group.
   \STATE Initialize $P^o_{i,j} \leftarrow $ empty lists, for $i,j \in
    \{1\dots n \}, j < i$.
   \FOR{$1$ {\bfseries to} $m$}
	\STATE Choose a surrogate patient $\tilde{s} \in 1\dots S$.
	\STATE Choose a subset $\tilde{S}$ of $\{1\dots S\}\backslash\tilde{s}$
		of $S$ surrogate controls.
	\STATE Compute $\widetilde{\B{\Sigma}^\star}$ and
	    $\widetilde{\B{d\Sigma}^s}$ for $s \in \tilde{S}$ using
	    algorithm \ref{alg:group_model} on the surrogate controls. 
	\STATE Compute $\widetilde{\B{d\Sigma}^s}$ for $s = \tilde{s}$,
	using $\widetilde{\B{\Sigma}^\star}$ and eqn
	\ref{eqn:tangent_model}.
	\STATE For all $i,j$, append to $P^o_{i,j}$ the T test comparing 
	$\widetilde{\B{d\Sigma}^s}_{i,j}$ for $s \in \tilde{S}$ and 
	for $s = \tilde{s}$.
   \ENDFOR
   \STATE Compute $\widehat{\B{\Sigma}^\star}$ and
       $\widehat{\B{d\Sigma}^s}$ for $s \in \tilde{S}$ using
       algorithm \ref{alg:group_model} on the complete control group.
   \STATE Compute $\widehat{\B{d\Sigma}^k}$,
   using $\widehat{\B{\Sigma}^\star}$ and eqn
   \ref{eqn:tangent_model}.
   \STATE For all $i,j$, compute $t_{i,j}$ the T test comparing 
    $\widehat{\B{d\Sigma}^s_{i,j}}$ for $s \in \tilde{S}$ and 
    for $s = \tilde{s}$.
   \STATE $p_{i,j} = 1 - \text{quantile}(\; t_{i,j} \;\text{in}
    \; P^o_{i,j})$
\end{algorithmic}
\end{algorithm}

\section{Algorithm evaluation on simulated data}
\vspace*{-1ex}%

\begin{figure}[tb]
\hspace*{-.01\linewidth}%
\includegraphics[width=.32\linewidth]{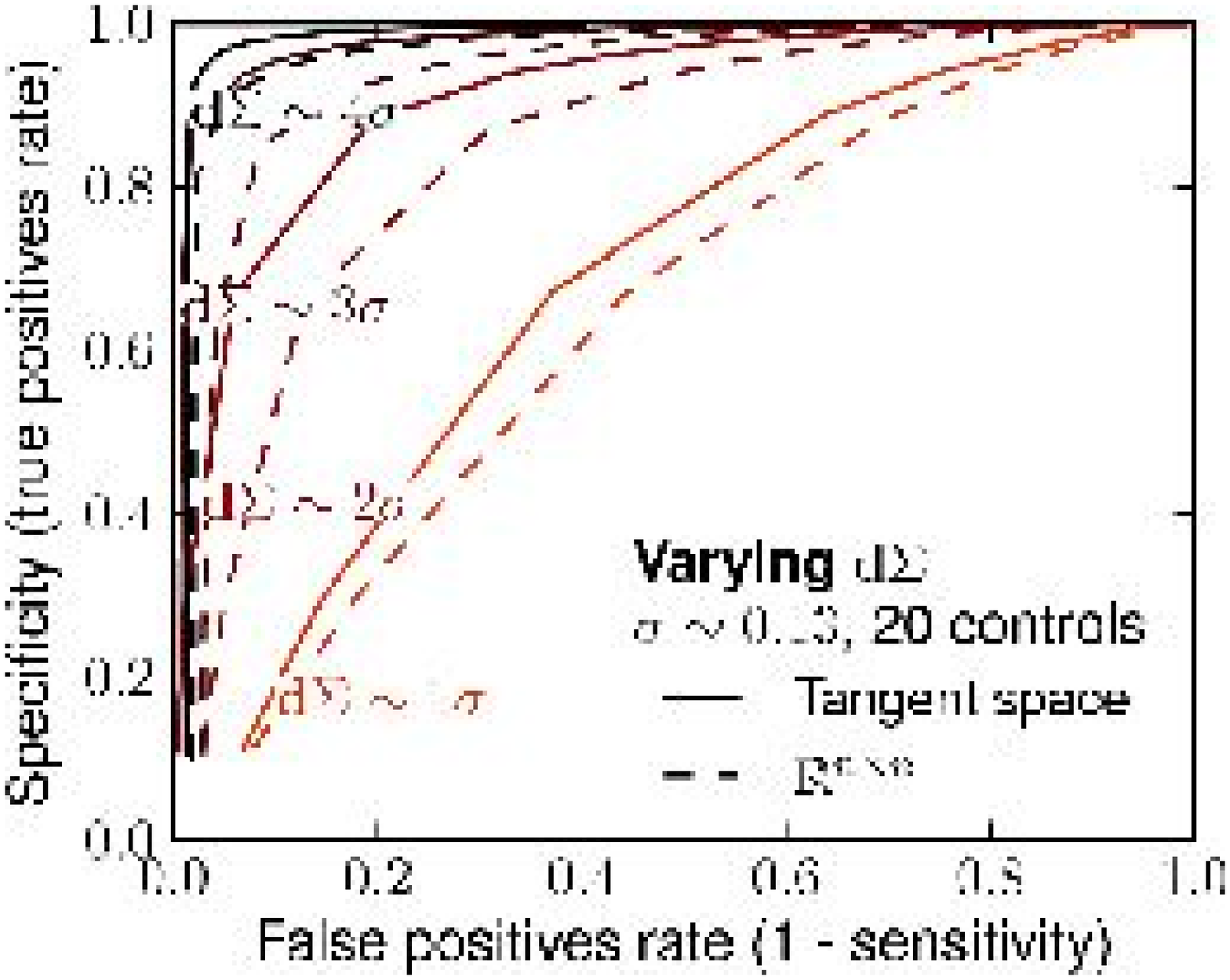}%
\llap{\bfseries\sffamily{\raisebox{18ex}{\scriptsize (a)}~~~}}%
\hfill%
\includegraphics[width=.32\linewidth]{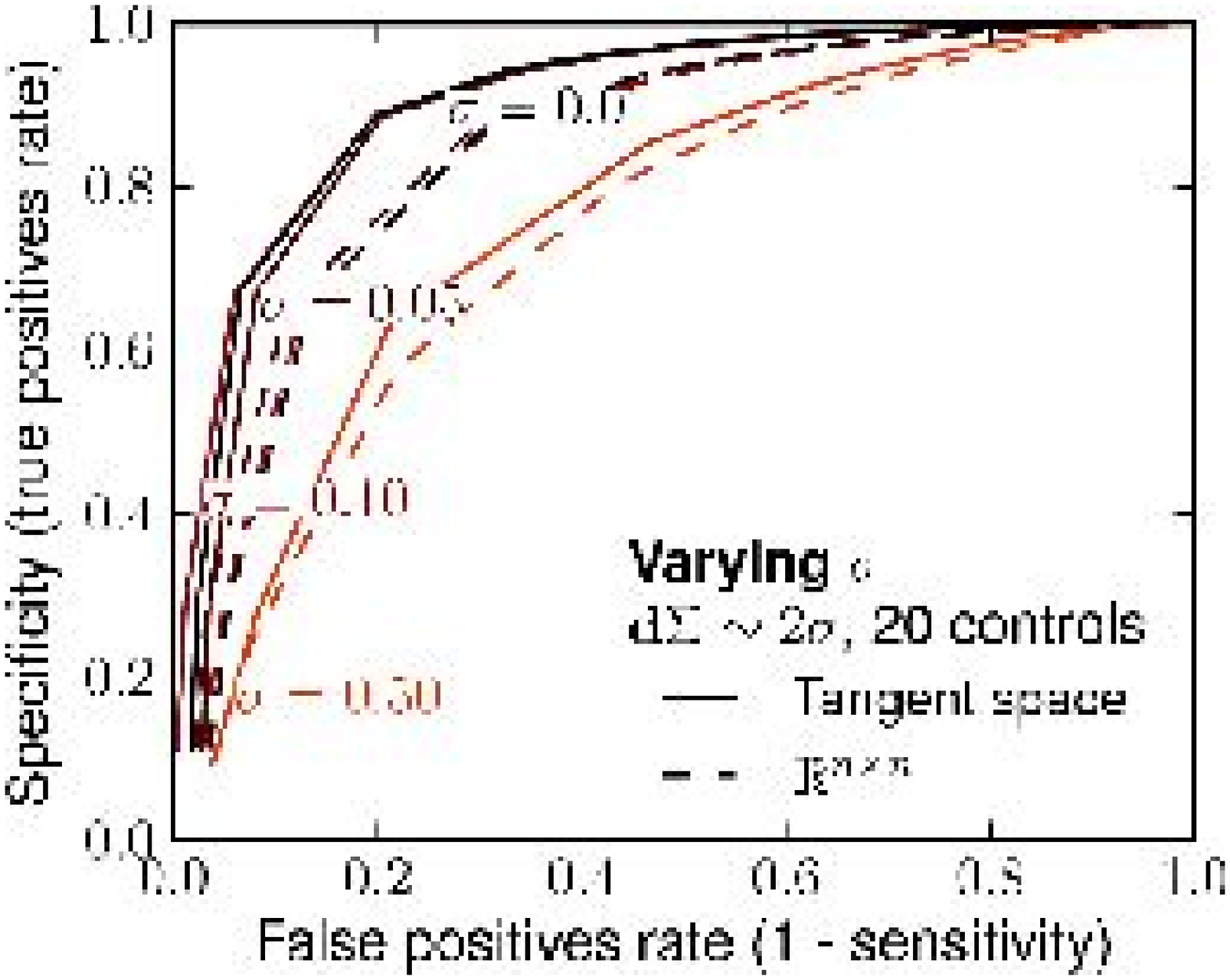}%
\llap{\bfseries\sffamily{\raisebox{18ex}{\scriptsize (b)}~~~}}%
\hfill%
\includegraphics[width=.32\linewidth]{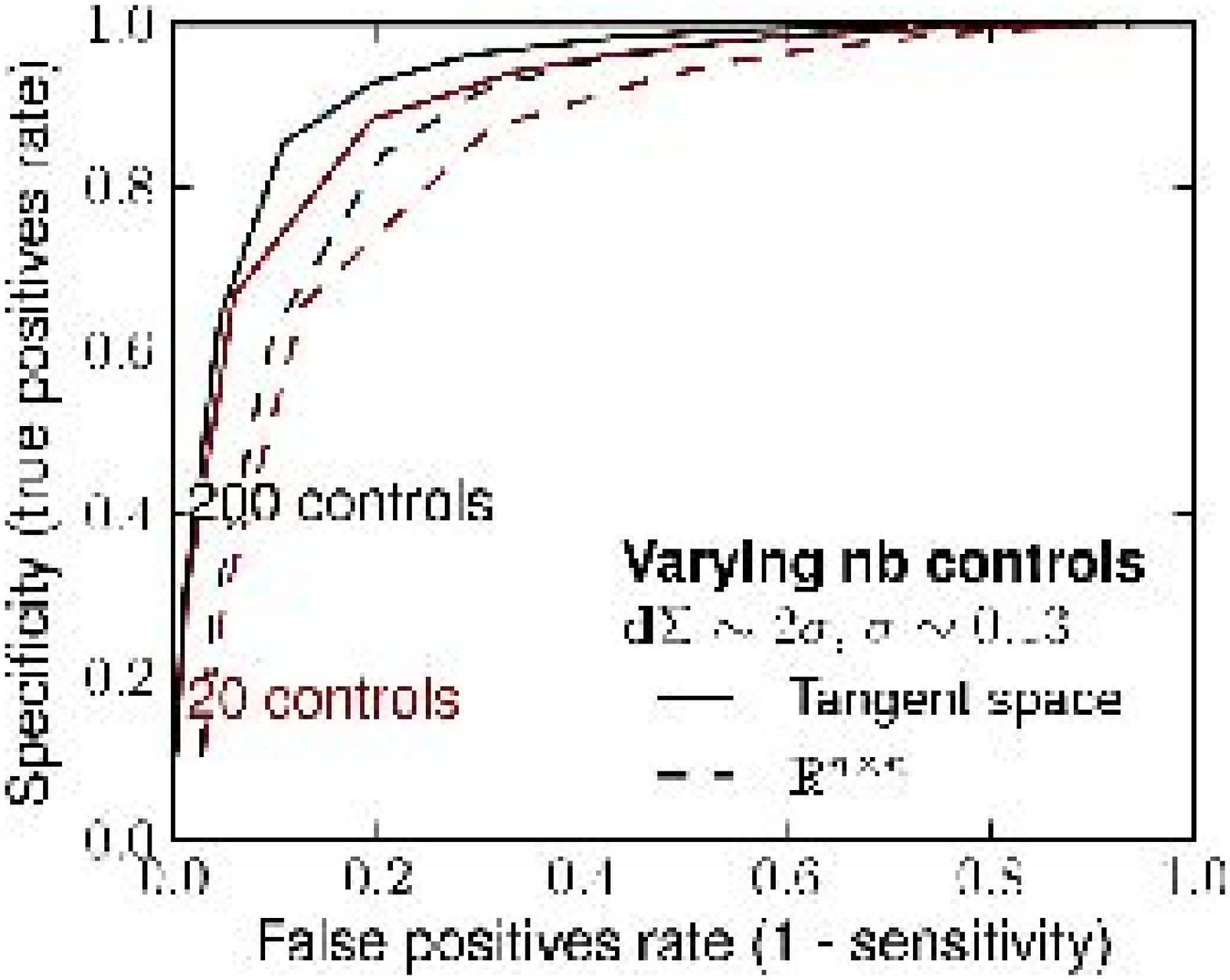}%
\llap{\bfseries\sffamily{\raisebox{18ex}{\scriptsize (c)}~~~}}%
\hspace*{-.01\linewidth}%
\vspace*{-1.5ex}%
\caption{ROC curves on data simulated according to the variability model given
by eqn \ref{eqn:tangent_model}.
{\bfseries\sffamily (a)} for different values of patient differences
$d\Sigma$. {\bfseries\sffamily (b)} for different values of control
variability $\sigma$. {\bfseries\sffamily (c)} for different number of
controls.
    \label{fig:roc}
}
\end{figure}

Algorithm \ref{alg:coefficient-level} relies on approximations of the
exact problem for coefficient-level detection of differences. In order to
quantify the performance of this detection, we study Receiver Operator
Characteristic (ROC) on simulated data: we draw a population of control
covariances using eq. (\ref{eqn:tangent_model}) with the parameters of
$\text{Vec}(\B{d\Sigma})$ normally distributed with deviation $\sigma$.
For simulated patients, we add differences of amplitude $d\Sigma$ to a
few coefficients ($\sim 20$) of this variability noise. For
$\B{\Sigma}^\star$ and $\sigma$, we use the parameters estimated on real
data (section \ref{sec:application}). We investigate the performance of
algorithm \ref{alg:coefficient-level} to recover these differences for a
variety of parameters. We observe good recovery for $d\Sigma > \sigma$
(Fig \ref{fig:roc}), and find that the comparison in the tangent space
(eq. \ref{eqn:tangent_model}) outperforms a comparison in $\mathbb{R}^{n
\times n}$ (eq. \ref{eqn:flat_model}).

\section{Application to post-stroke connectivity reorganization}
\vspace*{-1ex}%

\label{sec:application}

Standard clinical scores, such as the NIHSS, as well as fMRI studies can
be used to assess the consequences of cerebral strokes, but they test
specific cognitive functions and have little sensitivity to higher-order
cognitive malfunctions. Resting-state functional-connectivity is thus a 
valuable tool to study post-stroke reorganization. We apply our model to
stroke patients.

\paragraph{Resting-state fMRI dataset}
After giving informed consent, ischemic-stroke patients, as well as
age-matched healthy controls, underwent MRI scanning. Subjects with
existing neurology, psychiatry, or vascular pathologies were excluded
from the study. 10 patients and 20 controls were scanned during a
resting-state task: subjects were given no other task than to stay awake
but keep their eyes closed. 2 sessions of 10 minutes of fMRI data were
acquired on a Siemens 3T Trio scanner (245 EPI volumes, TR=2.46\,s, 41
slices interlaced, isotropic 3\,mm voxels). After slice-timing, motion
correction, and inter-subject normalization using SPM5, 33 ROIs were
defined in the main resting-state networks by intersecting a segmentation
of the gray matter with correlation maps from seeds selected from the
literature. For each subject, the BOLD time series corresponding to the
regions were extracted and orthogonalized with respect to confound time
series: time courses of the white matter and the cerebro-spinal fluid,
and movement regressors estimated by the motion-correction algorithm.
Covariance modeling was performed on the resulting 33 time series.

\begin{figure}[tb]
    \includegraphics[height=.25\linewidth]{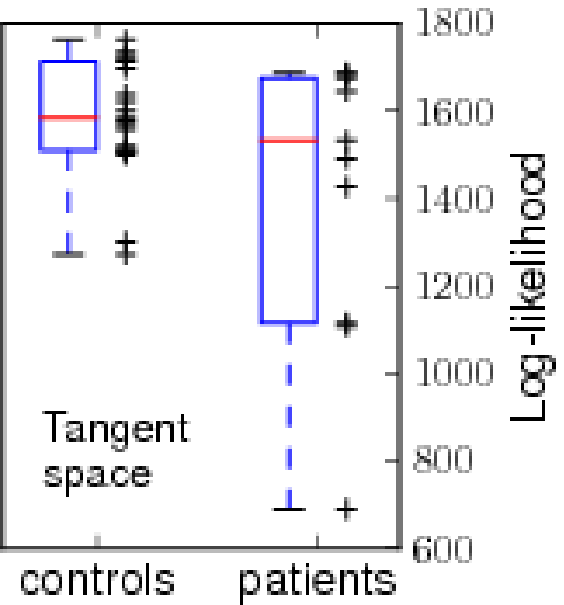}%
    \llap{\bfseries\sffamily \small (a)}%
    \hfill%
    \includegraphics[height=.25\linewidth]{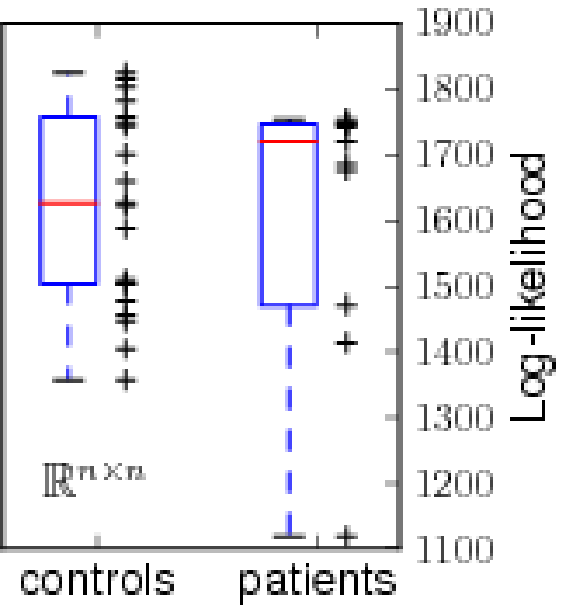}%
    \llap{\bfseries\sffamily \small (b)}%
    \hfill%
    \includegraphics[height=.25\linewidth]{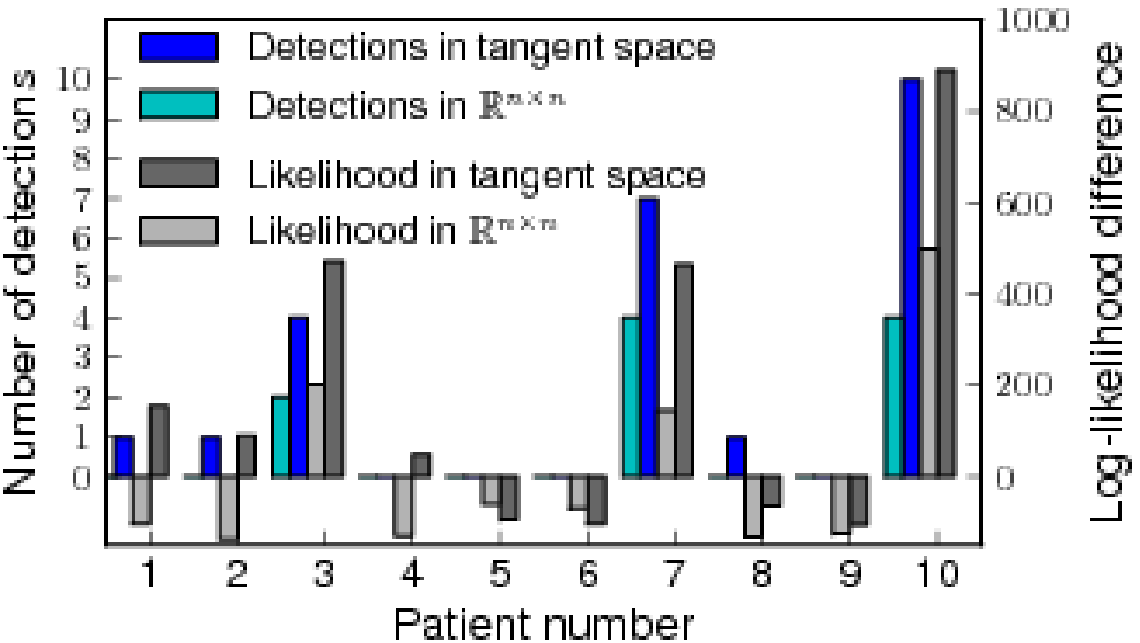}
    \llap{\bfseries\sffamily \small (c)~~~~}%
\vspace*{-2ex}%
\caption{
    {\bfseries\sffamily (a)} Likelihood of the controls and the patients in
    the model parametrized in the tangent space.
    {\bfseries\sffamily (b)} Likelihood of the controls and the patients in
    the model parametrized in $\mathbb{R}^{n \times n}$.
    {\bfseries\sffamily (c)} Number of coefficients detected as
    significantly different from the control group per patient, for the
    model parametrized in the tangent space, as well as in
    $\mathbb{R}^{n \times n}$.
    \label{fig:model_scores}
}
\end{figure}

\paragraph{Separating patients from controls with the matrix-variate
covariance model.}
To measure the discriminative power of the matrix-variate model
introduced in section \ref{sec:matrix_variate_model}, we test the
likelihood of patient data in a model learned on controls. Specifically,
we evaluate by {\sl leave one out} the likelihood of each control in the
model learned on the other controls. We compare this value to the average
likelihood of patients in the 20 models obtained by {\sl leave one out}.
We perform this comparison both using the group model isotropic on the
tangent space (eq. \ref{eqn:tangent_model}), and the group model
isotropic in $\mathbb{R}^{n \times n}$ (eq. \ref{eqn:flat_model}). We
find that the model in the tangent space separates better patients from
controls (Fig \ref{fig:model_scores}).

\paragraph{Detected connection differences}
We apply algorithm \ref{alg:coefficient-level} to detect the significant
coefficient-level differences for each subject. We compare to a similar
univariate procedure applied to the parametrization in $\mathbb{R}^{n
\times n}$ given by eq. (\ref{eqn:flat_model}), rather than the
tangent space. We find that coefficient-level analysis detects more
differences between ROI pairs when applied on the tangent-space
parametrization (Fig \ref{fig:model_scores}c). 

\begin{figure}[tb]
\vspace*{-3.5ex}%
    \hspace*{-1ex}%
    \includegraphics[width=.5\linewidth]{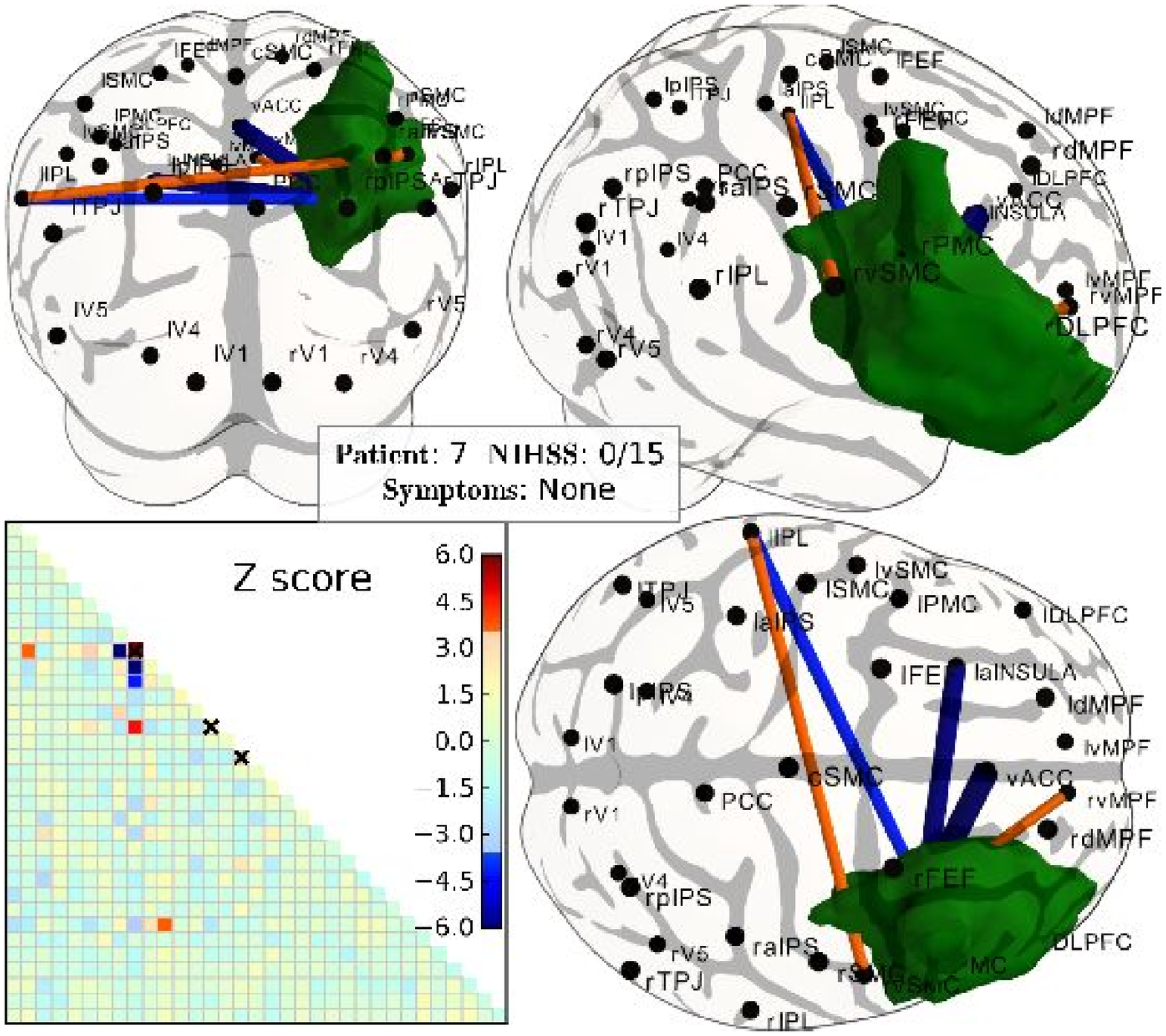}%
    \llap{\bfseries\sffamily \small (a)}%
    \hspace*{2ex}%
    \includegraphics[width=.5\linewidth]{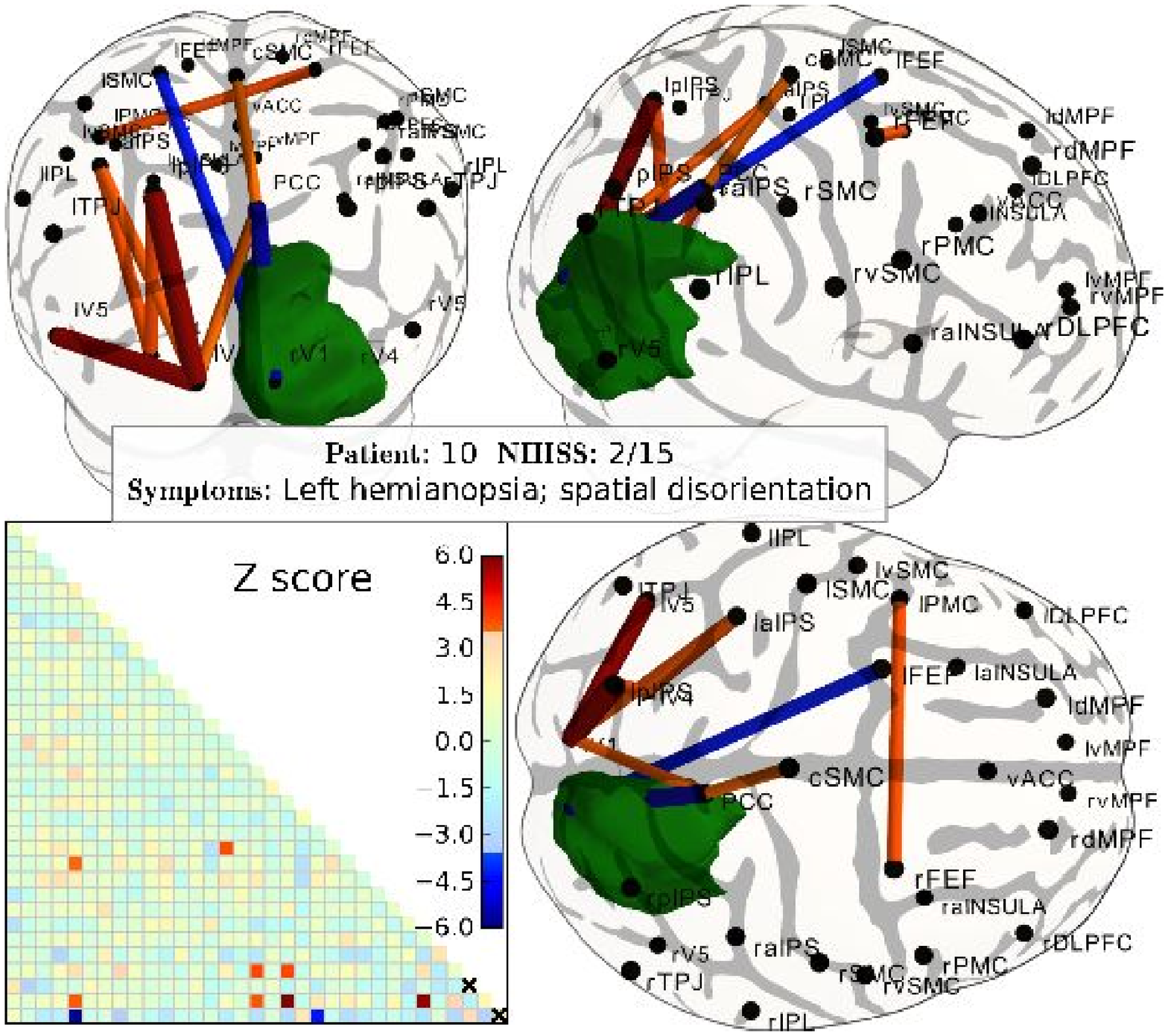}%
    \llap{\bfseries\sffamily \small (b)}%
\vspace*{-2ex}%
\caption{
Significant differences on two patients ($p < 0.05$ uncorrected),
represented as connections between regions: increased connectivity
appears in red, and decreased in blue. The lesion, manually segmented
from anatomical images, is represented in green. ROIs fully covered by a
lesion are marked with a black cross on the correlation matrix.
\label{fig:subject_graphs}
}
\end{figure}

\section{Discussion}
\vspace*{-1ex}%

\paragraph{Interpretation of the tangent space}
Projecting on the space tangent to the group mean corresponds to applying
a whitening matrix $\B{\Sigma^\star}^{-\frac{1}{2}}$ learned on the group
(eq. \ref{eqn:tangent_model}) that converts the Gaussian process
described by the group covariance to an independent and identically
distributed (iid) process. In other words, the {\sl coloring} of the time
series common to the group is canceled out to compare subjects on iid
coefficients on the correlation matrix.

\paragraph{Probing neurological processes}
For certain subjects, both procedures fail to detect a single connection
that makes a significant difference. Indeed, the variability of
resting-state activity in the control group induces some variability in
the projection to the tangent space. For patients with small lesions,
this variability is larger than the univariate differences. 
On the other hand, for patients with important lesions, the functional
connectivity analysis reveals profound differences in the correlation
structure that reflect functional reorganization. Some express a direct
consequence of the lesion, for example when the gray matter in one of two
ROIs has been damaged by the lesion, as can be seen on Fig
\ref{fig:subject_graphs}a. Others reflect functional reorganization. For
instance, patient 10 has a right visual cortex damaged by a focal lesion,
but the analysis shows increased connectivity in his left visual cortex
(Fig \ref{fig:subject_graphs}b). Functional connectivity analysis thus
reveals modifications that go beyond the direct anatomical consequences
of the lesion.

\section{Conclusion}
\vspace*{-1ex}%

We have presented a matrix-variate probabilistic model for covariances,
and have shown that it can be expressed as a random effect model on a
particular parametrization of the covariance matrix. The ability to draw
conclusions on the connectivity between pairs of regions is important
because it is a natural representation of the problem. We applied this
model to the comparison of functional brain connectivity between
subjects. We were able to detect significant differences in functional
connectivity between a single stroke patient and 20 controls.
A controlled detection of network-wide functional-connectivity
differences between subjects opens the door to new markers of brain
diseases as well as new insights in neuroscience, as functional
connectivity can probe phenomena that are challenging to access via
stimuli-driven studies.

\bibliographystyle{my_splncs}
\bibliography{restingstate}

\end{document}